\documentclass[noeprint,noshowpacs,longbibliography,nopreprintnumbers,twocolumn,prx,showpacs,superscriptaddress,amsmath,amssymb,citeautoscript,10pt,aps]{revtex4-2}
\usepackage{graphicx}
\usepackage{dcolumn}
\usepackage{color}
\usepackage{bm}
\usepackage[hidelinks]{hyperref}
\hypersetup{ colorlinks,citecolor=blue,filecolor=blue,linkcolor=blue,urlcolor=blue}

\usepackage{bbm}
\usepackage{graphics}
\usepackage{epsfig,color}
\usepackage[normalem]{ulem} 
\usepackage{soul}

\graphicspath{{Figures/}}

\newcommand{\be}{\begin{equation}}
\newcommand{\ee}{\end{equation}}
\newcommand{\bea}{\begin{eqnarray}}
\newcommand{\eea}{\end{eqnarray}}

\begin{document}

\title{Information propagation in one-dimensional \texorpdfstring{XY-$\Gamma$}{XY-Gamma} chains}

\author{Sasan Kheiri}
\affiliation{Department of Physics, University of Guilan, 41335-1914, Rasht, Iran}
\author{Hadi Cheraghi}
\email[e-mail: ]{h.cheraghi1986@gmail.com}
\affiliation{Institute of Physics, Maria Curie-Sk\l{}odowska University, 20-031 Lublin, Poland}
\affiliation{Computational Physics Laboratory, Physics Unit, Faculty of Engineering and Natural Sciences, Tampere
University, Tampere FI-33014, Finland}
\affiliation{Helsinki Institute of Physics, University of Helsinki FI-00014, Finland}
\author{Saeed Mahdavifar}
\affiliation{Department of Physics, University of Guilan, 41335-1914, Rasht, Iran}
\author{Nicholas Sedlmayr}
\affiliation{Institute of Physics, Maria Curie-Sk\l{}odowska University, 20-031 Lublin, Poland}

\begin{abstract}
The bond-dependent Kitaev model offers a playground in which one can search for quantum spin liquids. In these Kitaev materials, a symmetric off-diagonal $\Gamma$ term emerges, hosting a number of remarkable features, which has been particularly challenging to fully understand. One primary question that arises after recognizing a new phase is how information will spread in it. Out-of-time-ordered commutators and entanglement entropy describe processes whereby information about the initial condition of a unitarily evolving system propagates over the system.  A possible way to investigate dynamics in such systems is by considering one-dimensional models. We here investigate the one-dimensional spin-1/2 XY model in a transverse field with a $\Gamma$ interaction with periodic boundary conditions imposed. We will show that the $\Gamma$ interaction constructs an asymmetric ``light-cone'' with different butterfly velocities. In addition, it leads to faster information propagation in the spiral phase and slower propagation in the ferromagnetic and paramagnetic phases. Interestingly, we observe a pronounced effect in the entanglement entropy, explicitly showing up as a two-stage linear growth in time as fast/slow then slow/fast for quenches originating from the spiral phase. We hope our work paves the way for studying more about the spreading of information in one-dimensional Kitaev materials which can in turn help to discover unknown aspects of higher dimensional models.

\end{abstract}

\maketitle


\section{Introduction}\label{sec:int}

Ongoing discoveries in transition metal compounds have led to an enrichment of novel phases, including spin-orbit-entangled electronic phases \cite{Binz2009g,Yu2010g}. In this context the quantum spin liquid was introduced, in which the elementary excitations are fractionalized charge-neutral particles~\cite{Shirane1987g,Coldea2001g,Leon2010g}. This subject was begun by the seminal work of Jackeli and Khaliullin~\cite{Jackeli2009g} which introduced a realistic method for realizing the spin-1/2 Kitaev model, i.e.~an exactly solvable model on a two dimensional honeycomb lattice. Afterwards the potential for applications in quantum computers~\cite{Nayak2008g,Semeghini2021g} caused an intense growth in the study of Kitaev quantum spin liquid materials, such as the layered compounds $\rm{\alpha-RuCl_3}$ \cite{Plumb2014g, Banerjee2016g} and $\rm{A_2IrO_3}$ (A=Na, Li)~\cite{Hwan2015g, Williams2016g}, which exhibit magnetic order at low temperatures. However, in real materials additional spin interactions are inevitably present due to the lattice symmetries, suggesting the existence of an off-diagonal exchange between nearest neighbors, referred to as the $\Gamma$-interaction \cite{Rau2014g}. This term is believed to have a dominant effect over the Heisenberg term and has emerged as another source of frustration where its interplay with other interactions constructs a variety of complex orders~\cite{Modic2014g, Rousochatzakis2018g, Gordon2019g,Yang2020g,Nocera2020g}.

The phenomenology becomes much richer in a non-equilibrium setting where questions about information spreading arise. While most efforts on Kitaev quantum spin liquid models focus on the quantum critical lines, it is still unclear how information scrambles or entanglement entropy grows following a perturbation. The obstacle here is that there are strict difficulties in studying strongly correlated two dimensional systems for both analytical and numerical treatments. On the other hand, one dimensional systems are often easier to analyze and simulate, making them a favorable choice in many scenarios, even when exact or controllable approaches are not available. In this regard, the effects of the $\Gamma$ interaction on one dimensional spin-1/2 models have garnered a lot of interest~\cite{Nocera2020g, Agrapidis2018g, Liu2020g, Jize2021g, Herringer2022g, Xue2022g}. The model under investigation here has three equilibrium phases: ferromagnetic (FM), paramagnetic (PM), and a spiral pitched phase.

While considering probes of quantum chaos, out-of-time-order commutators (OTOCs)~\cite{Maldacena2016}  were found to be beneficial quantitative tools for characterizing scrambling. Scrambling in quantum systems is a process that describes how local information spreads and becomes inaccessible at later times. It generically characterizes the delocalization of quantum information after time evolution in many-body systems. The Lieb-Robinson bound~\cite{Lieb1972} provides an upper limit on the speed of information propagation in lattice systems with short-range interactions, bounded within a ``light-cone'' from the local dynamics, leading in turn to entanglement ``area laws''~\cite{Hastings2007g, Eisert2010g}, topological order~\cite{Bravyi2006g, Bravyi2010g}, and the decay of correlations~\cite{Koma2006g}. In this setting, the OTOC typically grows as $C(r,t) \propto e^{\lambda_L (t-r/{v_b})}$, where $\lambda_L$ and $v_b$ are referred to, respectively, as quantum analogs of the Lyapunov exponent~\cite{Maldacena2016} and the butterfly velocity~\cite{Bruno2006g, Roberts2016}. This spreading can occur at exponentially slow \cite{McGinley2019b} or fast \cite{Belyansky2020} rates. With the feasibility of observation in experiments \cite{Swingle2016,Garttner2017,Li2017,Lewis-Swan2019,Nie2020,Blok2021,Sundar2022}, OTOCs have attracted a lot of interest in physics across many different fields \cite{Larkin1969,Swingle2017, Buijsman2017, Patel2017, Dora2017, Dora2017a, Heyl2018, Chenu2018, Dag2019, Riddell2019, Sur2022a, Orito2022, Bin2023, Martyna2023g}. However, some questions emerge when the $\Gamma$ interaction is present; whether the signaling speed remains below the maximum group velocity. 

It is well established that a deep understanding of the dynamics of the coherence in a quantum system can be obtained by considering the behavior of entanglement entropy following a quantum quench~\cite{Pasquale2005g, DeChiara2006g} which provides one window on information spreading. The entanglement entropy originates from entanglement between the subsystem and its complement. 
Time-evolution typically generates correlations between subsystems as time goes on, resulting in an irreversible growth in the entanglement entropy~\cite{Kim2013g, Nahum2017}. The particular way in which the entanglement entropy grows in time is closely associated with the nature of the system. In integrable models linear growth is observed~\cite{Pasquale2005g}, in non-integrable models there is a linear growth which is much faster than the energy diffusion~\cite{Kim2013g}, in disordered systems following a quench  there is a logarithmic growth  from an unentangled initial state \cite{Maksym2013g} and a logarithmic logarithmic growth \cite{ Zsolt2012,Andraschko2016}, in many-body localized systems there is an unbounded logarithmic  growth \cite{Jens2012g}, and for long-range interacting spin systems a slow logarithmic growth is seen~\cite{Lerose2020g}. These characteristics enable one to distinguish quantum phases via the dynamics of the entanglement entropy. This in turn leads to the natural question of which aspects of the entanglement growth of the XY model undergo changes in the presence of the $\Gamma$ interaction.

Our results reveal that the $\Gamma$ interaction can induce an asymmetric light-cone in the dynamics of the OTOC. 
This behaviour also marks a difference between the spiral phase and the FM and PM phases. In the spiral phase the butterfly velocity for the positive separations is always larger than the maximum group velocity, whereas for the remaining two phases the opposite situation holds. One could quickly conclude that the $\Gamma$ interaction slows the speed of information propagation in the FM and PM phases. However, it actually increases the destruction of the information in the spiral phase. Interestingly, the $\Gamma$ interaction causes wave front changes under the influence of changing temperature, and as a result the parameters related to describing the OTOC behavior for early and long times will also change for different temperatures. 
In addition, we demonstrate that the dynamical behaviour of the entanglement entropy at zero temperature shows that control of the initial growth rate is possible via the $\Gamma$ interaction. Furthermore, quenches started from the spiral phase provide a two-stage growth which can be used as a signal to recognize this phase. We will also  discuss the value of the central change on the critical phase lines as well as within the phases.

This paper is organized as follows. In section II we present how one can calculate the OTOC and entanglement entropy. An introduction to the model and its critical lines can be found in section III. In section IV, we formulate how the OTOC and entanglement entropy are obtained for this model. Results and discussion are located in section V. Finally in section VI we conclude.


\section{OTOCs and entanglement entropy}\label{sec:scrambling}
\subsection{OTOCs}
The spreading of local perturbations 
is considered as one measure of information propagation in quantum systems, for which out-of-time-ordered commutators (OTOCs) are a central quantity, introduced as two-time correlation functions in which operators are not chronologically ordered in time. Typically one operator is fixed at a time $0$ and the other evolves from  $0$ to a time $t$. Let us consider two unitary operators $W_j$ and $V_{j+r}$ describing local perturbations to a lattice model at sites $j$ and $j+r$, respectively. The OTOCs are defined as the average of the squared commutator~\cite{Swingle2016,Roberts2016}, i.e.~as
\begin{eqnarray}
C(r,t)=\frac{1}{2}\left\langle\left[W_j(t),V_{j+r}(0)\right]^\dagger\left[W_j(t),V_{j+r}(0)\right]\right\rangle\,,
\end{eqnarray}
where for a given Hamiltonian $H$, the time evolution of $W_j$ is given by $W_j(t)=e^{i Ht}W_j(0)e^{-iHt}$. This means that as the operator $W_j(t)$ evolves in time a correlation develops with the perturbation at $V_{j+r}(0)$ as the operator ``spreads''. In the following the operators $W_j$ and $V_{j+r}$ are both also Hermitian allowing us to rewrite $C(r,t)$ as $C(r,t)=1-{\rm{Re}}[F(r,t)]$ in which $F(r,t)$ is the out-of-time ordered correlator
\begin{eqnarray}
F(r,t)=\left\langle W_j(t)V_{j+r}W_j(t)V_{j+r}\right\rangle\,.
\end{eqnarray}
It is conventional that if $C(r,t)$ vanishes or $F(r,t)$ goes to a large value in the long-time limit, the system signals the absence of scrambling, that is, no information has travelled from the site $j$ to $j+r$ in time $t$. The average $\langle {\cal O} \rangle={\rm Tr}(e^{-\beta H} {\cal O})/{\rm Tr}(e^{-\beta H})$ takes over the thermal ensemble with $\beta = 1/T$, the inverse temperature with the Boltzmann constant $k_B=1$. These quantities can detect the spread of quantum information beyond quantum correlations, in particular in quantum chaos where they signal growth bounded by a thermal Lyapunov exponent~\cite{Maldacena2016}.

The $\Gamma$ interaction breaks the mirror symmetry, hence, the only case that we can investigate under the Jordan-Wigner transformation applied to periodic boundary conditions is $C_{zz}(r,t)$ with $W_j(t)=\sigma_j^z(t)$ and $V_{j+r}=\sigma_{j+r}^z$. This restricts us from studying other cases, $\sigma_j^{x,y}$, since their calculations are based on the existence of the mirror symmetry and employment of the ``double trick''~\cite{Lin2018}. 
Near the wavefront of the spreading operators, integrable quantum systems unveil an exponential increase with time given by the Hausdorff-Baker-Campbell formula. This conjectured universal form describes the ballistic broadening of the OTOC given by~\cite{Lin2018,Xu2020a,Shenglong2019}
\begin{equation}\label{fitting}
C(r,t)\sim e^{-\lambda_L (r/{v_b}-t)^{1+d}t^{-d}}\,.
\end{equation}
The shape of the wave front is controlled by a single parameter $d$, associated with the growth rate $\lambda_L$, i.e.~the Lyapunov exponent, and the butterfly velocity $v_b$. There are some suggestions for $d$, including $d=1$ for a random circuit model~\cite{2018Khemani}, $d=1/2$ for a non-interacting translation invariant model~\cite{Xu2020a}, and $d=0$ for a Sachdev–Ye–Kitaev model~\cite{1993Subir}. We use equation \eqref{fitting} to access an estimate for the butterfly velocities on both sides of the light cone as $v_b^R$ and $v_b^L$, which refer to the right and left butterfly velocities respectively. 

\subsection{Entanglement entropy}

We aim to study the entanglement entropy after a global quench in our system at zero temperature ~\cite{Ferenc2009}. For a composite system with a Hilbert space ${\cal H} = {\cal H}_A \otimes {\cal H}_B$ in a pure quantum many-body state $ \rho  = | \Psi_0\rangle \langle \Psi_0|$, the entanglement entropy between subsystems $A$ and $B$ can be quantified by $S_{A/B}=-{\rm Tr}(\rho_{A/B} \ln\rho_{A/B})$ where the reduced density matrices are $\rho_{A/B}={\rm Tr}_{B/A} (\rho)$ ~\cite{Peschel2003g}. We focus on the case where $|\Psi_0\rangle$ is the ground state of our Hamiltonian. In this situation, the system is initially prepared in the ground state of the Hamiltonian. At $t = 0$, one parameter of the system is suddenly changed from its initial value to a final value and then the system evolves with the final Hamiltonian. In general, the entanglement entropy of a finite block $A$ of $l_a$ sites in an infinite system of free spinless fermions can be computed by~\cite{Peschel2003g}
\begin{eqnarray}
S_l(t)=-\sum\limits_{x = 1}^{2l_a} \lambda_x \ln(\lambda_x),
\end{eqnarray} 
where $\lambda_x$ are the eigenvalues of the $2l_a \times 2l_a$ correlation matrix $\cal M$:
\begin{eqnarray}
{\cal M} = \left( {\begin{array}{*{20}{c}}
\Theta &{\rm T}\\
{{\rm T}^\dag }&{\rm R}
\end{array}} \right).
\end{eqnarray}
$\Theta$, ${\rm T}$ and ${\rm R}$ are $l_a \times l_a$  matrices built from two-point correlation functions $\Theta _{nm} = \langle c_n^\dag {c_m}\rangle$, ${\rm T}_{nm} =\langle {c_n^\dag c_m^\dag }\rangle$, and $R_{nm} =\delta_{nm}-\Theta_{mn}$.
Here $c_n^\dag$ ($c_n$) is the fermionic creation (annihilation) operator.
It has been illustrated that for one dimensional integrable models, the entanglement entropy does indeed spread ballistically~\cite{Pasquale2005g}, growing like the boundary area of the subsystem $A$, and not like its volume, which is known as the ``area law''. Noncritical ground states of short-range interacting spin chains with a finite correlation length have a constant entanglement entropy. At a quantum critical point, when the subsystem $A$ is a finite interval of length $L/2$, the entanglement entropy slightly violates the area law by a logarithmic correction as, $S_{L/2}(L) = (c_{eff}/3) \log(L)+b$, where $c_{eff}$ is the effective central charge~\cite{2004Calabrese1,Hastings2007g} and $b$ is a non-universal constant.


\section{\texorpdfstring{XY-$\Gamma$}{XY-Gamma} model}\label{sec:model}
The XY model is one of the benchmark integrable models which is equivalent to the Kitaev chain. We consider a 1D spin-1/2 XY chain in a transverse field in the presence of a generalized $\Gamma$ interaction. The Hamiltonian $H=H_{XY}+H_{\Gamma} $ reads
\begin{eqnarray}
H_{XY} &=& J\sum\limits_{n = 1}^{L}[ (\frac{1 + \delta}{2}) \sigma_n^x\sigma_{n + 1}^x + (\frac{1 - \delta}{2})\sigma_n^y\sigma_{n + 1}^y]  \nonumber\\
&&+ h\sum\limits_{n = 1}^{L}\sigma_n^z\,,  \textrm{ and} \nonumber\\
H_{\Gamma} &=&\Gamma \sum\limits_{n = 1}^L \left( \sigma _n^x\sigma _{n + 1}^y + \gamma \sigma _n^y\sigma _{n + 1}^x \right)\,,
\end{eqnarray}
where $\sigma_n^{x,y,z}$ are the usual Pauli matrices. Also, $J$, $\delta$, and $h$ are the antiferromagnetic coupling, the anisotropy parameter, and the strength of the uniform transverse field, respectively. In addition $\Gamma$ characterizes the amplitude of the off-diagonal exchange interactions while $\gamma$ denotes the relative coefficient of the off-diagonal exchange couplings. These parameters decide the phases and properties of this model. We impose periodic boundary conditions so that $\sigma_{L+1}=\sigma_1$ with $L$ the length of the spin chain.

\begin{figure}[t]
\centerline{\includegraphics[width=1.1\linewidth, height=0.23\textheight]{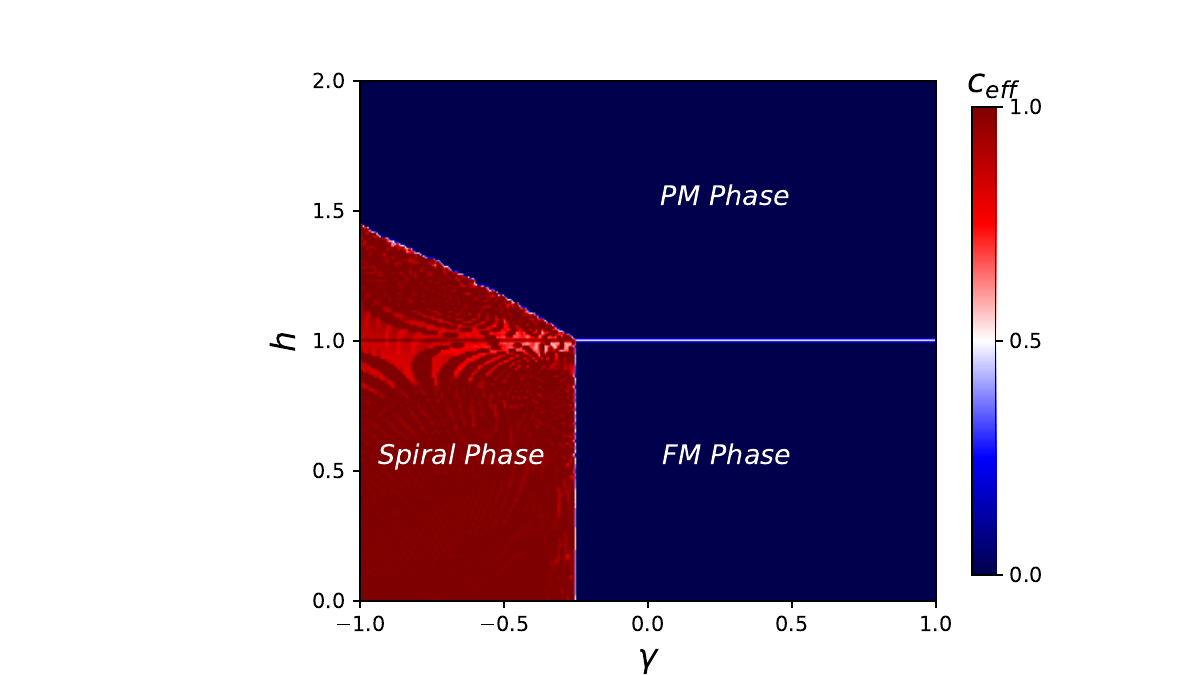}}
\caption{Effective central charge $c_{eff}$ versus $h$ and $\gamma$ obtained by fitting $S_{L/2}(L)$. A gapless conformal field theory with $c_{eff}=1/2$ is visible on the critical line between the FM and PM phases while on the other critical lines the central charge is zero. In addition, within the FM and PM phases, it is zero while within the spiral phase, its value is one. Here and also in other figures we fix $J=1.0$, $\delta=0.6$, and $\Gamma=0.6$. Note that, as we see in the spiral phase, there are some fluctuations, especially around $h=1$ which arise from finite size effects. Thus, by increasing the system size, tending to the thermodynamic limit $L \to \infty $, they will vanish.}
\label{Fig1}
\end{figure}

The Hamiltonian is analytically solved by a Jordan-Wigner transformation $\sigma_n^+=\exp[i\pi \sum_{m<n}c_m^\dag c_m]c_n$ and $\sigma_n^z=2c_n^\dag c_n-1$ \cite{Lieb1961g}, followed by a Fourier transformation $c_n=(1/\sqrt{L})\sum_k \exp[ikn]c_k$ where the possible values of $k$ should be given for a fixed value of $L$, and finally Bogoliubov transformations~\cite{Xue2022g} given by
\begin{eqnarray}
c_k &=& \cos (\Phi _k)\eta _k - \sin (\Phi _k)e^{i\theta _k}\eta _{-k}^\dag\,\textrm{, and}\nonumber\\
c_{ - k}^\dag  &=&  \cos (\Phi _k)\eta _{-k} ^\dag+ \sin (\Phi _k)e^{-i\theta _k}\eta _k\,.
\end{eqnarray}
Following these transformations the diagonalized Hamiltonian reads
\begin{eqnarray}
H =\sum\limits_k \varepsilon _k (\eta _k^\dag \eta _k - 1/2)
\end{eqnarray}
with $\varepsilon _k= P_k + \sqrt {A_k^2 + B_k^2 + Q_k^2}$, where
\begin{eqnarray}
A_k&=&2[J\cos(k) +h]\,,\nonumber\\
B_k&=& 2J\delta \sin (k)\,, \nonumber\\
P_k&=&2\Gamma(\gamma  - 1)\sin (k)\,,\textrm{ and} \nonumber\\
Q_k&=&2\Gamma(\gamma + 1)\sin (k)\,.
\end{eqnarray}
According to the conditions $\theta _{-k}=\theta _k$ and $\Phi _{-k}=-\Phi _k$ the Bogoliobov angles will be
\begin{eqnarray}
\tan (2\theta _k) &=& \frac{2B_kQ_k}{Q_k^2-B_k^2}\,,\textrm{ and} \nonumber\\
\tan (2\Phi _k) &=&\rm{sgn}(k)~ \frac{\sqrt {B_k^2 + Q_k^2}}{A_k} \,,
\end{eqnarray}
where $\rm{sgn}(k)$ is the sign function defined as 1 for $k\ge0$ and $-1$ for $k<0$. The $\Gamma$ interaction accords this model several nontrivial quantum phase transitions and properties. The ground-state phase diagram of the model consists of three phases: the gapped ferromagnetic and paramagnetic phases separated by $h_{c_1}=1$ for $\gamma>\delta^2/(4\Gamma^2)$, and the gapless spiral phase characterized by a quasi-long-range order separated from the FM phase by $\gamma_{c_1}=\delta^2/(4\Gamma^2)$ for $h\le1$ and from the PM phase by $h_{c_2}=\sqrt{1-\delta^2-4\Gamma^2\gamma}$ for $\gamma<\delta^2/(4\Gamma^2)$.
With this rich phase diagram it is interesting to study how information spreads in this model in these different phases.

Throughout this paper we set $J=1$ as the energy scale and fix the parameters $\delta=0.6$ and $\Gamma=0.6$, which still allows us to reach all phases. Nothing qualitative depends on this choice, which simplifies the presentation of the results. Therefore only $h$ and $\gamma$ are left as free parameters for which the ground-state phase diagram is depicted in Fig.~\ref{Fig1}.


\section{Methods}

\subsection{OTOC}

We are interested in the OTOCs for the case when the perturbations $W$ and $V$ are given by single-site Pauli matrices such as $\sigma_j^z$. Consequently, one can write 
\begin{eqnarray}
F_{zz}(r,t) = \langle {\sigma _r^z(t)\sigma _0^z\sigma _r^z(t)\sigma _0^z} \rangle.
\end{eqnarray}
Since the model is exactly solvable by means of the Jordan-Wigner transformation, it is convenient to express the Pauli matrix by fermionic operators as $\sigma_j^z=-A_jB_j$, with $A_j=c_j^\dag + c_j^\dag$, $B_j=c_j^\dag - c_j^\dag$, where $c_j^\dag$ ($c_j$) is the fermionic creation (annihilation) operator.
Let us rewrite the Hamiltonian in $k$ space as $H=\sum_{k>0} H_k$  in the eigenbasis $\left\{ |0_k0_{ - k}\rangle ,| 1_k1_{ - k}\rangle ,| 1_k0_{ - k}\rangle ,| 0_k1_{ - k}\rangle \right\}$. This helps us to write the density of state at the time $t=0$ in the form
\begin{eqnarray}
\rho _k(t=0) =\frac{1}{\Omega_k} \left[ {\begin{array}{*{20}{c}}{d_{11}}&{d_{12}}&0&0\\
{d_{21}}&{d_{22}}&0&0\\0&0&{d_{33}}&0\\0&0&0&{d_{44}}\end{array}} \right]
\end{eqnarray}
with  $ \Omega _k = 2\left[ \cosh (\beta \Lambda _k^{(1)}) + \cosh (\beta P_k^{(1)}) \right]$ and
\begin{eqnarray}
d_{11/22} &=& \cosh (\beta \Lambda _k^{(1)}) \pm \cos(2\Phi_k^{(1)})\sinh (\beta \Lambda _k^{(1)})\,, \nonumber\\
d_{12/21}  &=& -e^{\mp i \theta_k^{(1)}} \sin(2\Phi_k^{(1)}) \sinh(\beta \Lambda _k^{(1)})\,,       \nonumber\\
d_{33/44} &=& e^{\mp \beta P_k^{(1)}}\,,    
\end{eqnarray}
where $\Lambda _k = \sqrt{A_k^2 + B_k^2+Q_k^2}$.

\begin{figure}[t]
\centerline{\includegraphics[width=0.9\linewidth, height=0.2\textheight]{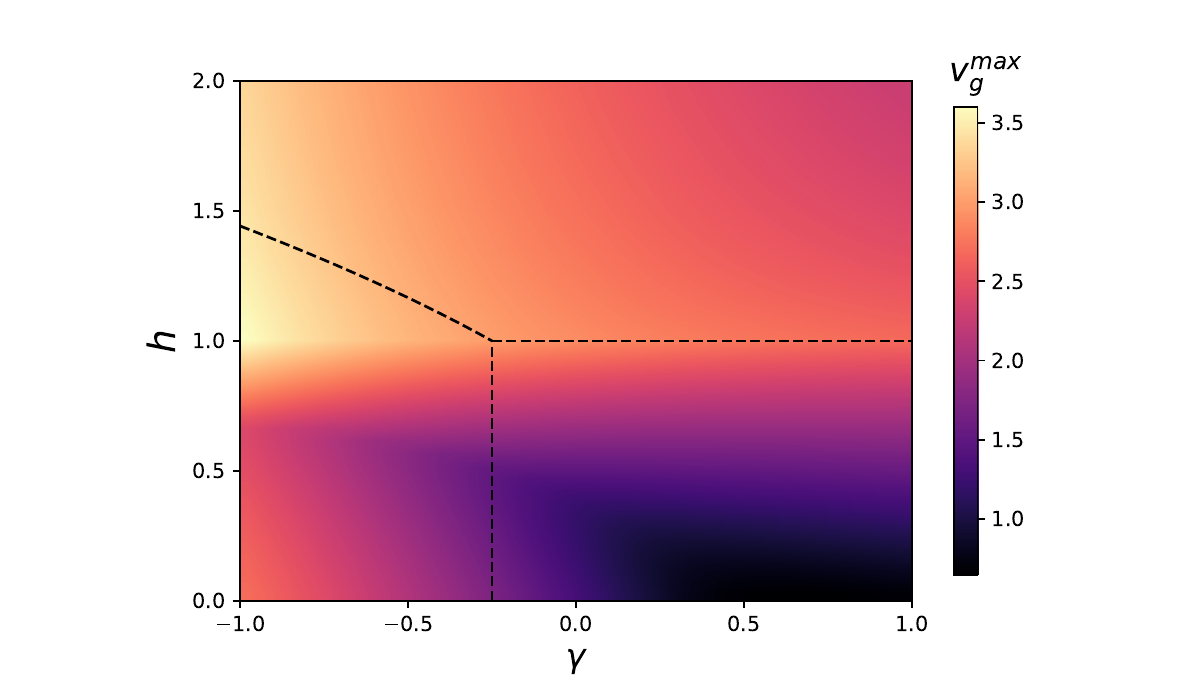}}
\caption{Density plot of the maximum of the group velocity versus $h$ and $\gamma$. The dashed black lines show the critical lines.}
\label{Fig2}
\end{figure}

On the other hand, the unitary time-evolution operator will drive
\begin{eqnarray}
U_k(t)=e^{ -2itJ\cos(k)}\left[ {\begin{array}{*{20}{c}}
{k_{11}}&{k_{12}}&0&0\\
{k_{21}}&{k_{22}}&0&0\\
0&0&{k_{33}}&0\\
0&0&0&{k_{44}}
\end{array}} \right]
\end{eqnarray}
with
\begin{eqnarray}
k_{11/22} &=&   \cos (t \Lambda _k^{(2)}) \pm i \cos(2\Phi_k^{(2)})\sin (t \Lambda _k^{(2)})\,, \nonumber\\
k_{12/21}  &=&  -i e^{\mp i \theta_k^{(2)}}\sin(2\Phi_k^{(2)}) \sin (t \Lambda _k^{(2)})\,,   \nonumber\\
k_{33/44} &=&  e^{\mp it P_k^{(2)}}\,.
\end{eqnarray}
The indices (1) and (2) refer to the pre- and post-quench Hamiltonians, respectively.
Note that in the basis, the fermionic operators read
\begin{eqnarray}
c_k = \left[ {\begin{array}{*{20}{c}}0&0&1&0\\0&0&0&0\\0&0&0&0\\0&1&0&0\end{array}} \right]~&;
~c_k^\dag  = \left[ {\begin{array}{*{20}{c}}0&0&0&0\\0&0&0&1\\1&0&0&0\\0&0&0&0\end{array}} \right]&
\nonumber\\
c_{ - k} = \left[ {\begin{array}{*{20}{c}}0&0&0&1\\0&0&0&0\\0&{ - 1}&0&0\\0&0&0&0\end{array}} \right] ~&;
~c_{ - k}^\dag  = \left[ {\begin{array}{*{20}{c}}0&0&0&0\\0&0&{ - 1}&0\\0&0&0&0\\1&0&0&0\end{array}} \right]&
\end{eqnarray}
and therefore their time-evolution can now be obtain from ${\cal O} _k(t) = U_k^ \dag(t) {\cal O}_k U_k(t)$. With these at hand, one can calculate the required time-dependent correlation functions, which are 
\begin{widetext}\begin{align}
\langle A_n(t)A_m \rangle  =&   \frac{1}{L}\sum\limits_k e^{ik(m-n)} \langle U_k^\dag(t) (c_k^\dag + c_{-k})U_k(t)(c_{-k}^\dag  + c_k) \rangle\,, \nonumber \\
\langle B_n(t)B_m \rangle  =& \ \frac{1}{L}\sum\limits_k e^{ik(m-n)}  \langle U_k^\dag(t)(c_k^\dag - c_{-k})U_k(t)(c_{-k}^\dag  - c_k) \rangle\,,  \nonumber \\
\langle A_n(t)B_m \rangle  =&\frac{1}{L}\sum\limits_k e^{ik(m-n)} \langle U_k^\dag(t)(c_k^\dag + c_{-k})U_k(t)(c_{-k}^\dag  - c_k) \rangle\,, \nonumber \\
\langle B_n(t)A_m \rangle  =&  \frac{1}{L}\sum\limits_k e^{ik(m-n)} \langle  U_k^\dag(t)(c_k^\dag - c_{-k})U_k(t)(c_{-k}^\dag  + c_k)\rangle\,.
\end{align}
\end{widetext}
$m$ and $n$ denote the position of operators in the spin chain.

\begin{figure*}[t]
\centerline{\includegraphics[width=0.9\linewidth]{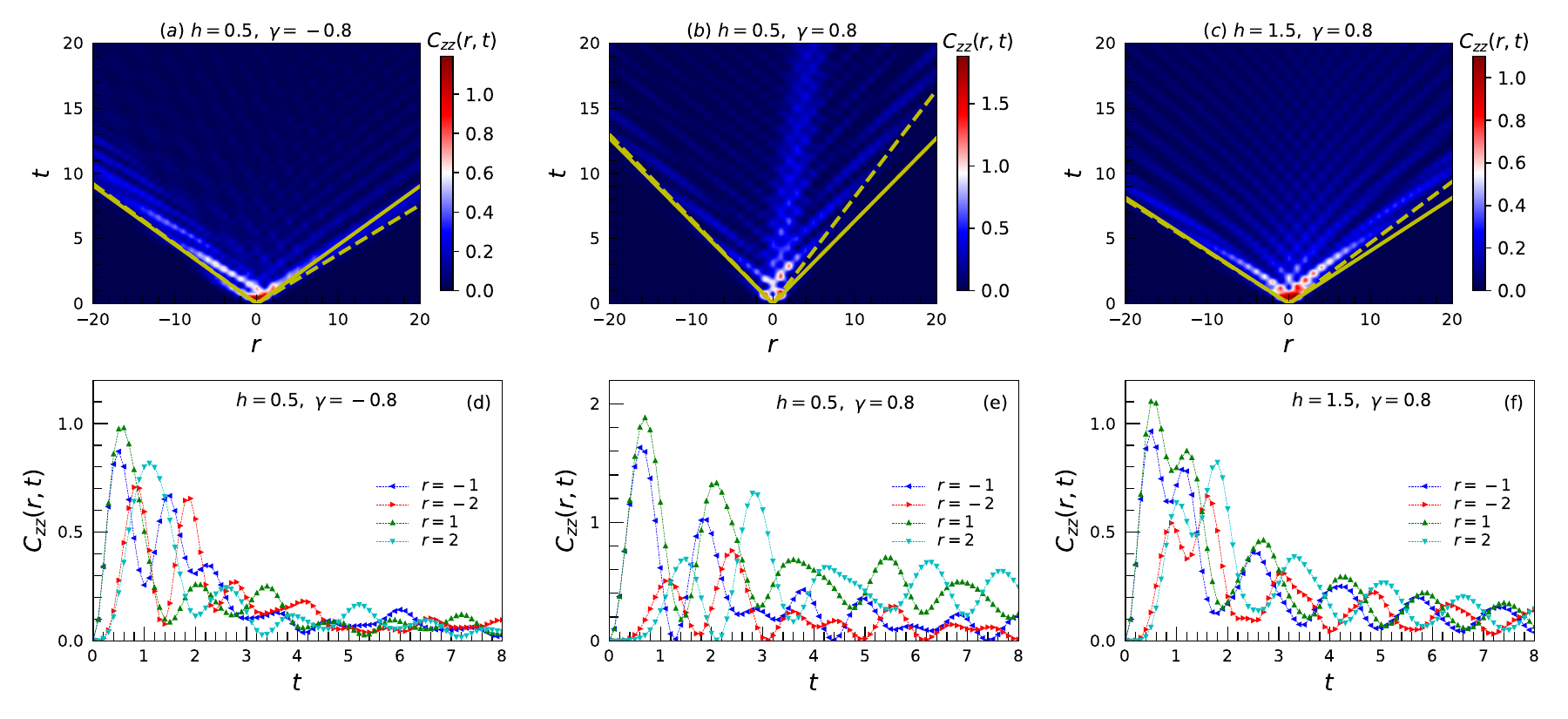}}
\caption{Density plot of $C_{zz}(r,t)$ versus separation, $r$, and time, $t$, for (a) $h=0.5$, $\gamma=-0.8$, (b) $h=0.5$, $\gamma=0.8$, and (c) $h=1.5$, $\gamma=0.8$. 
 The yellow lines are an aid to the eye for how fast the correlations spread, the dashed lines show the butterfly velocities and the solid lines are the maximum group velocities. The absolute values of the maximum group velocity $v_g^{max}$, the butterfly velocities of right $v_b^R$ and left $v_b^L$ are (a) $v_g^{max}\approx v_b^L \approx 2.208$, $v_b^R\approx 2.642$,  (b) $v_g^{max}\approx v_b^L \approx 1.574$, $v_b^R\approx 1.25$, and (c) $v_g^{max}\approx v_b^L \approx  2.453$, $v_b^R\approx 2.127$. In the lower panels $C_{zz}(r,t)$ is plotted at fixed separations $r = \pm 1,\pm 2$ versus time for (d) $h=0.5$, $\gamma=-0.8$, (e) $h=0.5$, $\gamma=0.8$, and (f) $h=1.5$, $\gamma=0.8$. Here the numerical simulations are done for $L = 100$, inverse temperature $\beta = 0$, and $J=1.0$, $\delta=0.6$, $\Gamma=0.6$.}
\label{Fig3}
\end{figure*}

\subsection{Entanglement entropy}
As previously mentioned, in this section we aim to study the entanglement entropy following global quenches where the initial state is a ground state. The ground state can be found from the condition that $\eta_{\pm k} |GS \rangle=0$ if $\varepsilon _{\pm k} \ge 0$ and $\eta_{\pm k}^\dag |GS \rangle=0$ if $\varepsilon _{\pm k} < 0$. In this respect, the ground state in general will be
\begin{align}
|GS\rangle  &= \prod\limits_{k \notin \cup({\varpi _+},{\varpi _-})} | 0_k,0_{ - k} \rangle \otimes \prod\limits_{k \in {\varpi _+}} {\eta _k^\dag | 0_k,0_{ - k}\rangle } \nonumber\\
&\otimes  \prod\limits_{k \in {\varpi _-}} {\eta _{ - k}^\dag | 0_k,0_{ - k} \rangle }
\end{align}
in which $\varpi_{\pm}$ denotes a $k$ range with $\varepsilon_{\pm k}<0$ where $\varpi_+=-\varpi_-=\varpi$. Here $|0_k,0_{ - k}\rangle$ is the vacuum of the Bogoliubov quasiparticles $\eta_{\pm k}|0_k,0_{ - k}\rangle=0$. We now can attain the time-dependent two-point correlation functions through
\begin{align} 
\Theta _{nm}  =& \frac{1}{L}\sum\limits_{k} \cos[k(m-n)] |v_k(t)|^2 \nonumber\\
&+ \frac{1}{L}\sum\limits_{k \in \varpi}  \left\{  \cos[k(m-n)](|u_k(t)|^2-|v_k(t)|^2)\right. \nonumber\\
&+ \left.  i\sin[k(m-n)]  \right\}  
\end{align}
and
\begin{align}
{\rm T}_{nm}  =&   \frac{i}{L}\sum\limits_{k} \sin[k(m-n)] u_k(t) v_k^*(t) \nonumber\\
&-  \frac{2i}{L}\sum\limits_{k \in \varpi} \sin[k(m-n)] u_k(t) v_k^*(t) \,.
\end{align}
We also have
\begin{align} 
v_k(t) =&  -e^{i\theta _k^{(1)}}\left\{ \sin (\Phi _k^{(1)})\cos (t\Lambda _k^{(2)}) \right.  \nonumber\\
&- i\sin (t\Lambda _k^{(2)}) \big[\sin (\Phi _k^{(1)})\cos (2\Phi _k^{(2)})  \nonumber\\
&\qquad- \left. \cos (\Phi _k^{(1)})\sin (2\Phi _k^{(2)})e^{i\Delta \theta _k} \big] \right\}  
\end{align}
and
\begin{align}
u_k(t) =& \cos (\Phi _k^{(1)})\cos (t\Lambda _k^{(2)})  \nonumber\\
&+ i\sin (t\Lambda _k^{(2)}) \big[ \cos (\Phi _k^{(1)})\cos (2\Phi _k^{(2)}) \nonumber\\
&+  \sin (\Phi _k^{(1)})\sin (2\Phi _k^{(2)})e^{ - i\Delta \theta _k}  \big] \,,
\end{align}
with $\Delta \theta _k=\theta _k^{(2)}-\theta _k^{(1)}$. 

\begin{figure*}[!t]
\centerline{\includegraphics[width=0.92\linewidth]{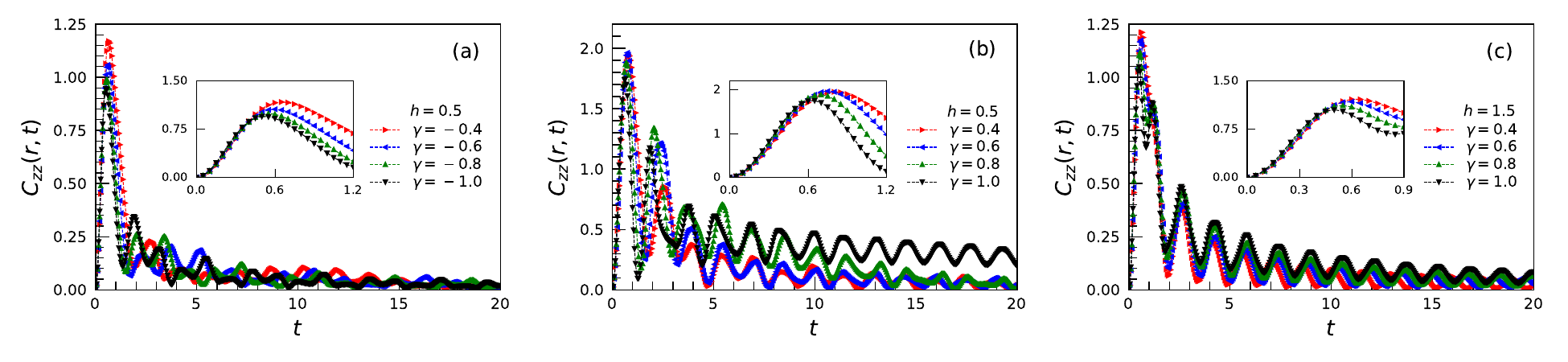}}
\caption{Dynamics of $C_{zz}(r,t)$ for $r=1$, $L=100$ and $\beta=0.0$ as (a) $h=0.5$, $\gamma=-0.4,-0.6,-0.8,-1.0$, (b) $h=0.5$, $\gamma=0.4,0.6,0.8,1.0$, and (c) $h=1.5$,  $\gamma=0.4,0.6,0.8,1.0$. The insets indicate the early time evolution.}
\label{Fig4}
\end{figure*}


\section{Results and discussion}\label{sec:ssh}

Following the calculations outlined in the previous section we can find the OTOCs and entanglement entropy for a range of scenarios. In this section we discuss the results of these calculations, first looking at the OTOCs.

\subsection{The OTOC}
We here focus on $C_{zz}(r,t)$ within the different phases as a function of the temperature. First we consider the case where there is no quench, i.e.~the initial density matrix is the ground state of the time evolving Hamiltonian. We have also investigated quenches in which the initial density matrix and the Hamiltonian belong to different phases. The results of this latter case can be found in appendix \ref{app:quenches}.

Before starting to study how the OTOC evolves in time, in Fig.~\ref{Fig2} we plot the maximum of the quasiparticle group velocity $v_g = \partial {\varepsilon _k}/\partial k$,
\begin{align}
v_g=&2\Gamma(\gamma-1)\cos(k)\\
&+\frac{2\sin(k)}{\Lambda_k}\left[(J^2(\delta^2 - 1)+\Gamma ^2(\gamma+1)^2)\cos(k)-Jh\right] \nonumber.
\end{align}
This is compared to the envelope on the OTOC function~\cite{Lieb1972}. Here we can immediately identify that the maximum group velocity is dependent on both $\gamma$ and $h$ when the rest of the parameters are fixed. In addition the largest value of $v_g^{max}$ is in the spiral phase around $h=1.0$.

In Fig.~\ref{Fig3} $C_{zz}(r,t)$ is plotted at infinite temperature $\beta=0.0$ for three different phases: (a) and (d) the spiral phase, (b) and (e) the FM phase, and (c) and (f) the PM phase. The results illustrate how the different phases affect the evolution of the OTOC and the spreading velocity of the butterfly effect. Here, we choose the system size $L=100$. The density plots explicitly indicate asymmetric  propagation in the system in all three phases which originates from the presence of the $\Gamma$ interaction. The values of the left butterfly velocities are very close to the values of the maximum group velocities. In contrast, for propagation to the right, this matching is absent. Consequently, the maximum group velocities clearly do not give any absolute bound. We note that an asymmetry in propagation has also been reported for the helical multiferroic chains around the ballistic wavefront  where the topologically nontrivial quantum phases allows for electric-field controlled anisotropic propagation~\cite{2021Sekania}. Additionally, asymmetric OTOC and light cones also emerge in  the non-equilibrium dynamics of Abelian anyons in a 1D system~\cite{2021Shun}.

Interestingly, in the spiral phase, Fig.~\ref{Fig3}(a), the right  butterfly velocity always has a bigger value than maximum group velocity while in the other two phases this is the opposite. This can be interpreted as a signatures of slow or fast operator spreading in these phases. Therefore, the results support that $v_b$ depends on the $\Gamma$ interaction. In addition, Fig.~\ref{Fig3}(b) indicates that the system in the FM phase reveals a narrower light cone compared to the other phases, with a slower spreading of the local operator which expresses slower information propagation.  In order to show the difference in propagation in the two directions, we also have drawn $C_{zz}(r,t)$ for $r=\pm 1, \pm 2$ in Fig.~\ref{Fig3}(d-f). The mismatch is clearly evident for example between $r=1$ and $r=-1$ for each case. Further, the value of $C_{zz}(r,t)$ for the first peak for the positive $r$ always is bigger than the corresponding negative one.

\begin{figure}[!t]
\centerline{\includegraphics[width=1.1\linewidth, height=0.47\textheight]{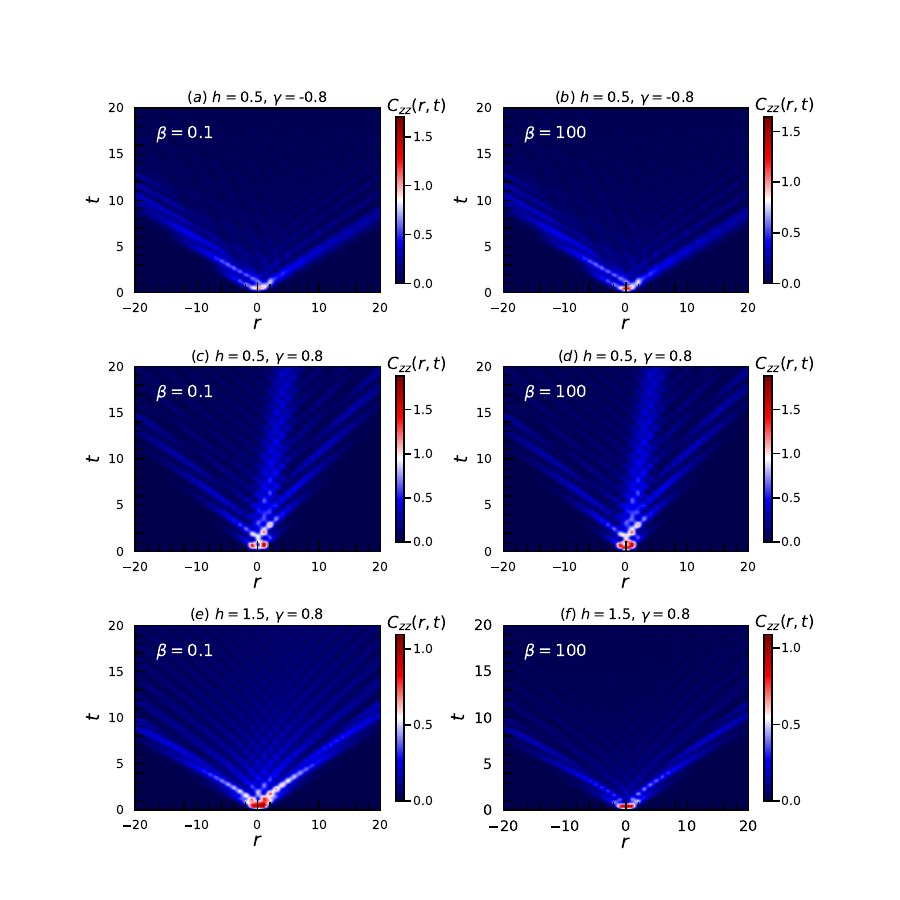}}
\caption{Density plot of $C_{zz}(r,t)$ versus separation, $r$, and time, $t$, for a chain with size $L=100$ as (a),(b) $h=0.5$, $\gamma=-0.8$, (c),(d) $h=0.5$, $\gamma=0.8$, and (e),(f) $h=1.5$, $\gamma=0.8$. The left, and right columns, respectively,  belong to $\beta=0.1$  and $\beta=100$. }
\label{Fig5}
\end{figure}

We have continued our study for $C_{zz}(r,t)$ by considering the impact of the $\Gamma$ interaction on the spread of the information. Here we focus on a fixed site, the case $r=1.0$. As is clear from Fig.~\ref{Fig4}, $C_{zz}(r,t)$ typically increases in a short time from zero to its maximum value and then decreases, vanishing at long times in an oscillating manner. In fact, by increasing the value of $\gamma$ in the FM (Fig.~\ref{Fig4}b) and PM (Fig.~\ref{Fig4}c) phases, a slower decay is found. This means that the $\gamma$ interaction prevents a quick loss of information in the system. However, this preservation of the information is very remarkable for the case $\gamma=1.0$ where the system stays in a symmetric situation. The opposite case emerges for the spiral phase in which the increase of the absolute value of $\gamma$ makes for a quicker decay (Fig.~\ref{Fig4}a). However,  we see that the OTOCs comprised of local operators show no sign of scrambling,
$\lim _{t \to \infty }C_{zz}(r,t) = 0$. The insets of Fig.~\ref{Fig4} also clearly display that the decay rate at short times after the first growth is strongly affected by an increase in $\gamma$, and the rate of decay will be higher. In App.~\ref{FittingLieb} we present an example of the fitting used to estimate the parameters $v_b$, $\lambda_L$, and $d$ for the different phases for $r=\pm 1$.

We have also considered the effect of different temperatures in the system. Results for $\beta=0.1,100$ are illustrated in Fig.~\ref{Fig5}.  As before, the data is for $\Gamma = 0.6$, and the calculations are done for a system size $L = 100$.
We see that the temperature has no effect on the shape of the light-cone. However, it can change the value of $C_{zz}$ in the PM phase, although in the FM and spiral phases this effect appears only in the short-time behavior at small separations. It is worth mentioning that for the case where $\Gamma=0.0$, it has been indicated that the temperature has a negligible effect on OTOCs with local operators~\cite{Lin2018}, except for the case $\gamma=0.0$, $h=1.0$~\cite{2020Bao}. Here we can conclude that a temperature-dependent description of the wavefront, especially in the PM phase, is still possible in the presence of the $\Gamma$ term. On the other hand, any ``universal'' description needs to be essentially temperature independent. For this reason, the growth of the decay rates at long times of $C_{zz}$ in our model do not follow the reported universalities~\cite{Lin2018}, c.f.~App.~\ref{FittingLieb}.

\begin{figure}[!t]
\centerline{\includegraphics[width=1.15\linewidth]{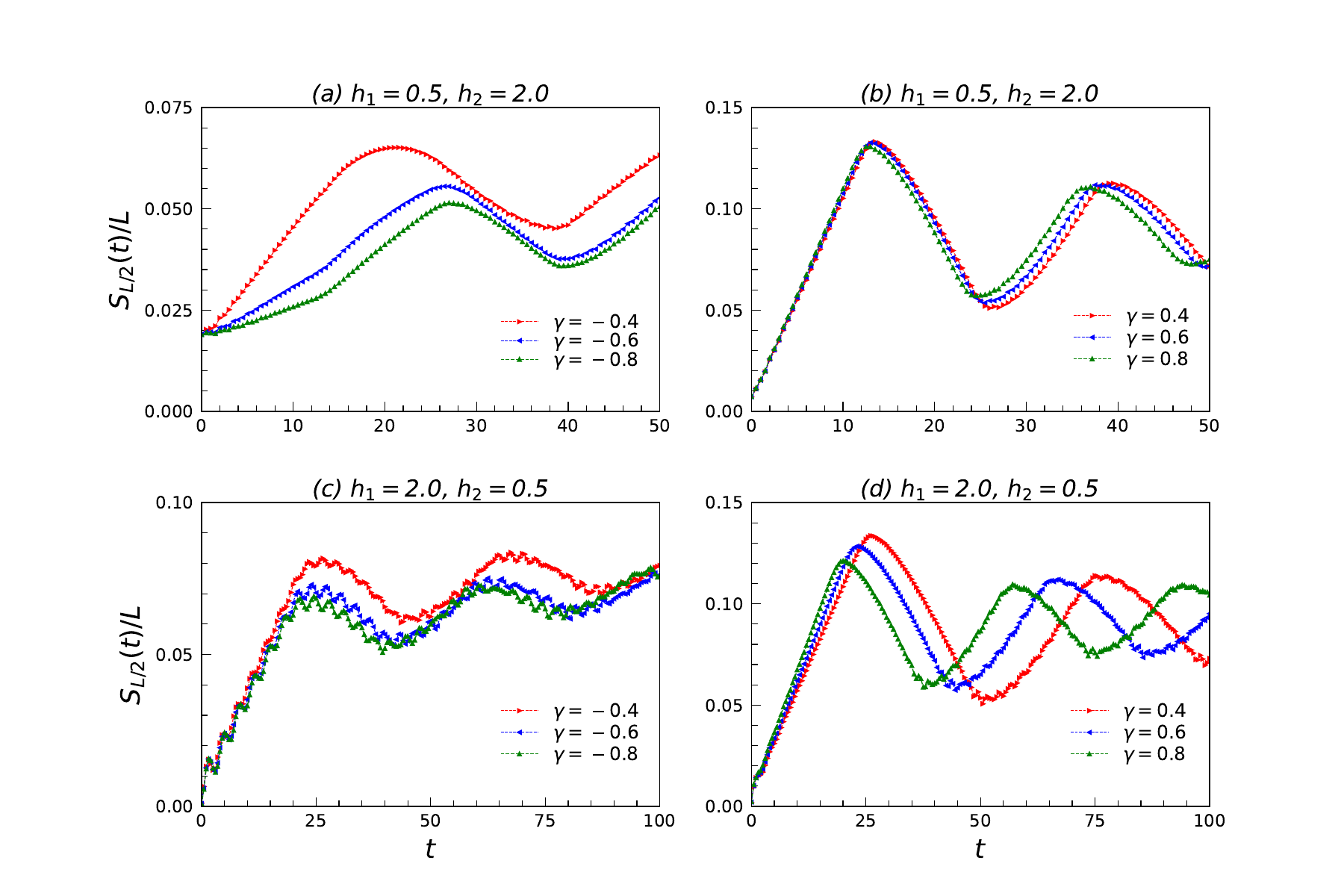}}
\caption{Dynamics of $S_{L/2}(t)/L$ for $L=100$ as (a) and (b) $h_1=0.5$, $h_2=2.0$, (c) and (d) $h_1=2.0$, $h_2=0.5$ for initial states in (a) the spiral phase, (b) the FM phase, (c) and (d) the PM phase.}
\label{Fig6}
\end{figure}
\begin{figure}[!t]
\centerline{\includegraphics[width=1.15\linewidth]{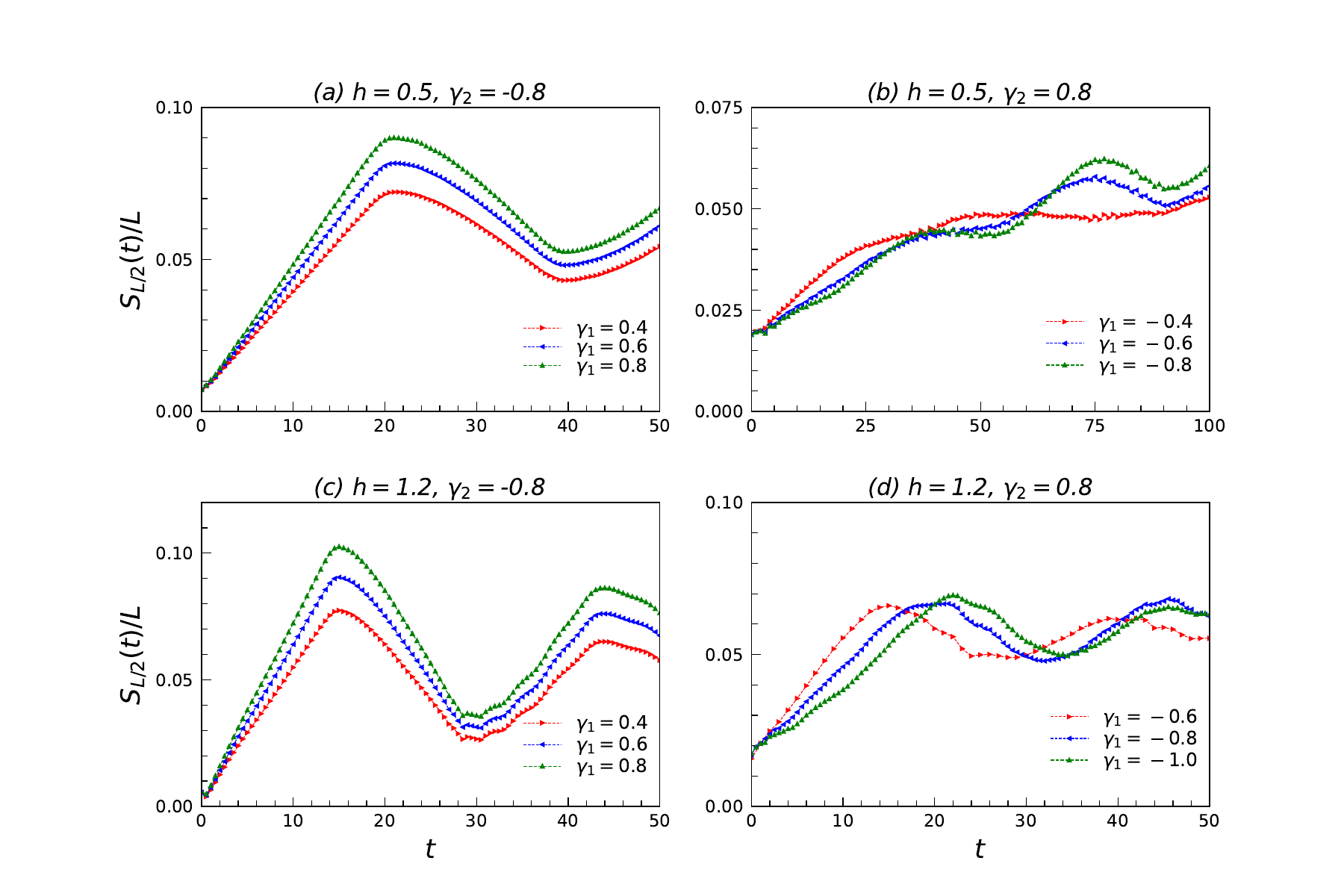}}
\caption{Dynamics of $S_{L/2}(t)/L$ for $L=100$ as (a) and (b) $h=0.5$, (c) and (d) $h=1.2$. The quenches are from (a) the FM (c) the PM phases to the spiral phase, and from the spiral phase to (b) the FM and (d) the PM phases. }
\label{Fig7}
\end{figure}


\subsection{Entanglement entropy}

Entanglement entropy, as a fundamental concept, provides a key route to understanding many-body quantum systems in and out of equilibrium. Indeed, it measures gross quantum mechanical correlations between different parts of a system. In equilibrium, the ground-state phase diagram of the model is depicted  against $h$ and $\gamma$ in Fig.~\ref{Fig1}. As viewed, on the critical line between the FM and PM phases, the central charge is $1/2$ while on other critical lines it has a zero value.
Intriguingly, within the spiral phase, the central charge is one which is consistent with the fact that the low-energy excitations of the gapless region belong to the same universality class as the Tomonaga-Luttinger liquid. In addition, within the FM and PM phases, $c_{eff}$ is zero, described by Ising-like excitations. However, some fluctuations are visible in the spiral phase caused by finite size effects. In the thermodynamic limit these fluctuations will vanish. To confirm our results, we also studied the entanglement entropy as a function of different system sizes and different subsystems at fixed system size. These results are in App.~\ref{Equilibrium}.  

\begin{figure}[t]
\centerline{\includegraphics[width=1.15\linewidth]{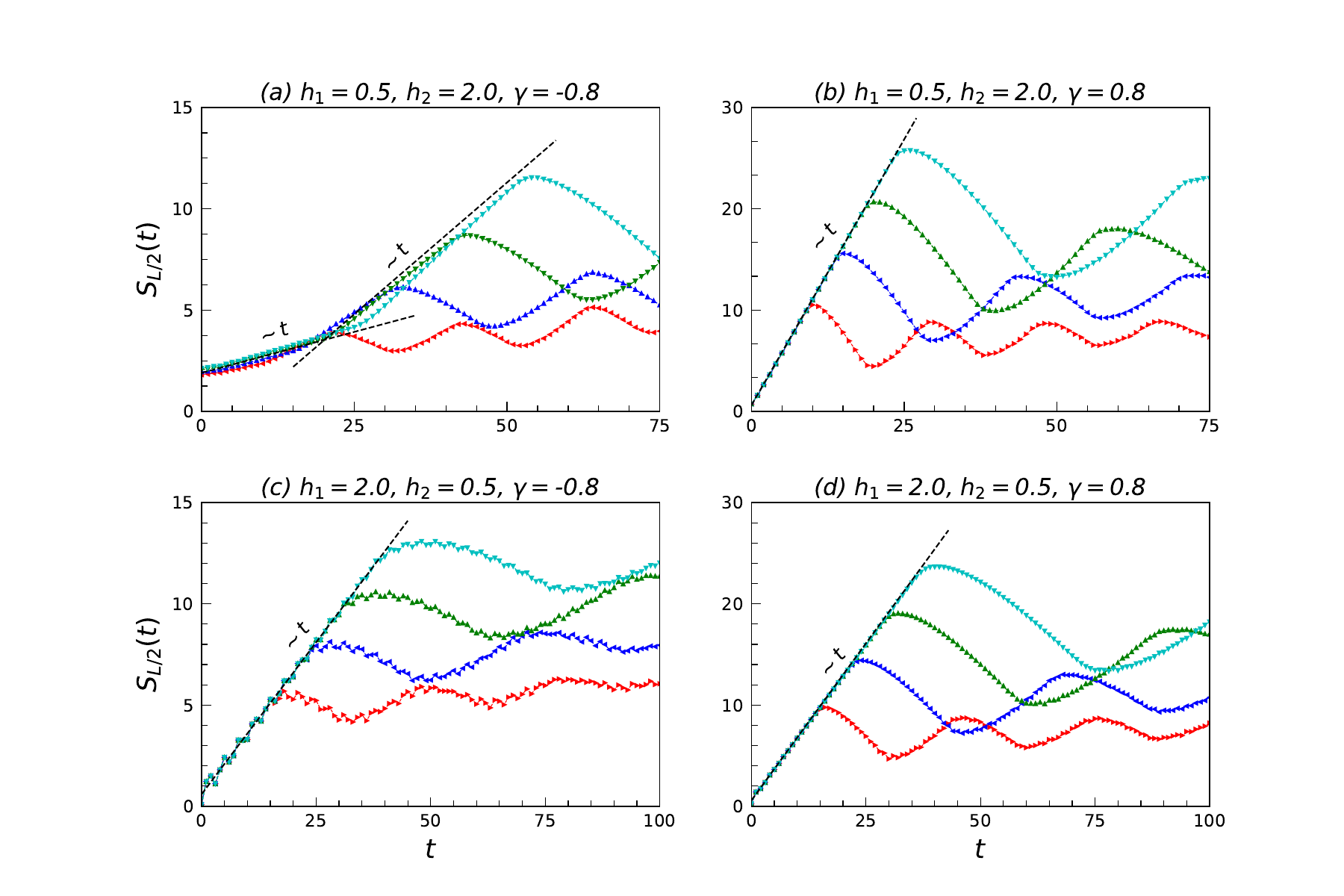}}
\caption{Dynamics of $S_{L/2}(t)$ for different system sizes as $L=80,120,160,200$, (from red to cyan) as (a) $h_1=0.5$, $h_2=2.0$, $\gamma=-0.8$, (b) $h_1=0.5$, $h_2=2.0$, $\gamma=0.8$, (c) $h_1=2.0$, $h_2=0.5$, $\gamma=-0.8$, and (d) $h_1=2.0$, $h_2=0.5$, $\gamma=0.8$. The black dashed lines are a guide for the eyes, representing the initial growth rate $S_{L/2}(t) \sim t$.}
\label{Fig8}
\end{figure}

\begin{figure}[t]
\centerline{\includegraphics[width=1.15\linewidth]{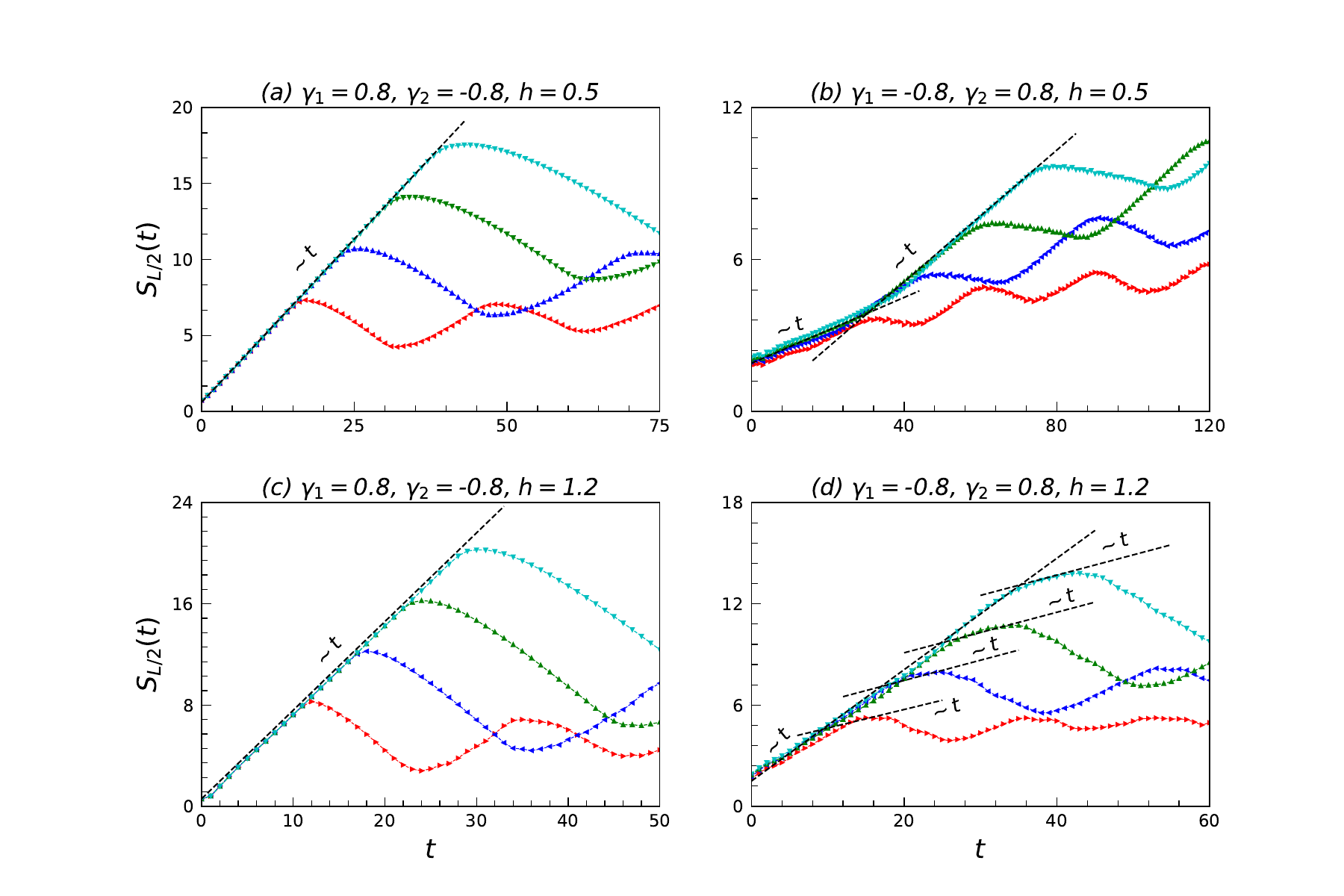}}
\caption{Dynamics of $S_{L/2}(t)$ for different system sizes as $L=80,120,160,200$, (from red to cyan) as (a) $\gamma_1=-0.8$, $\gamma_2=0.8$, $h=0.5$, (b) $\gamma_1=0.8$, $\gamma_2=-0.8$, $h=0.5$, (c) $\gamma_1=-0.8$, $\gamma_2=0.8$, $h=1.2$, and (d) $\gamma_1=0.8$, $\gamma_2=-0.8$, $h=1.2$. The black dashed lines are a guide for the eyes, representing the initial growth rate $S_{L/2}(t) \sim t$.}
\label{Fig9}
\end{figure}

Here we study the impact of the $\Gamma$ interaction using two strategies:  with quenches between different phases (see Figs.~\ref{Fig6} and \ref{Fig7}), and the entanglement entropy growth for a given quench under different system sizes (see Figs.~\ref{Fig8} and \ref{Fig9}).  As depicted, we consider quench protocols where at $t =0$ the state of the whole system is prepared as the ground state of the pre-quench Hamiltonian at zero temperature. Then, the post-quench Hamiltonian instantaneously drives the time evolution of the system.

In Fig.~\ref{Fig6}, we show the entanglement entropy for a system with size $L=100$ where quenches are done in the transverse field $h$ for constant values of $\gamma$. In this setting, the system is quenched from the spiral and the FM phases into the PM phase and vice versa. As we can see, except for the quench from the FM into the PM phase, Fig.~\ref{Fig6}(b) increasing the value of $\gamma$ reduces remarkably the value of the entanglement entropy.For the quench from from the FM into the PM phase a small reduction in the entanglement entropy is found. On the other hand, the initial growth rate shows a different behaviour. Quenches between the spiral and the PM phases uncover a decrease of the growth rate while in quenching between the FM and the PM phases, an increase of the growth rate emerges when increasing $\gamma$. However, the decrease of the growth rate is significant for the quench originating from the spiral phase, as is clear from Fig.~\ref{Fig6}(a).

In contrast, when quenches are caused by changing $\gamma$ with the transverse field constant, Fig.~\ref{Fig7}, regardless of the initial and final phases an increase in $\gamma$ causes an enhancement of the entanglement entropy at long times.  In particular, quenches starting from the spiral phase increase the initial growth rate while quenches from the other phases into the spiral phase reduce the initial growth rate. These imply that $\gamma$ can function as a quench control parameter which is able to control the initial growth rate in the system.

In the one dimensional XY model, the respective entanglement entropies are expected to rise linearly with time during unitary evolution~\cite{2008Maurizio}. Here, in Figs.~\ref{Fig8} and \ref{Fig9}, the time evolution of the entanglement entropy is shown for several system sizes $L=80,120,160,200$ for different quenches covering the phase diagram. Our main goal is to determine how the initial entanglement entropy grows with time. As is clearly displayed, the linear growth is visible in all quenches, $S_{L/2}(t) \sim t$. The highlight is that for quenches starting from the spiral phase, the entanglement entropy shows a two-step linear growth. This growth, depending on the phase which is quenched into, can be first slow and then fast or vice versa, and consequently could be used as a sign to detect the spiral phase. There is a linear time regime followed by nonlinear behaviour. Indeed, the ballistic growth continues up to a crossover time $t^*$ where it begins to saturate. In general, we see in our our numerical data that the crossover time does not always obey $t^*=L/(2v_g^{max})$. This happens because sometimes the modes with maximum velocity can carry less information than others~\cite{2008Maurizio,2017Najafi, 2020Hadi}.


\section{Conclusions}\label{sec:concl}

In order to shed light on the role that $\Gamma$ interactions play in the behaviour of higher dimensional systems we have, in this paper, considered an exactly solvable 1D spin-1/2 XY model in the presence of a transverse field and $\Gamma$ interaction. The ground-state phase diagram of our model consists of three different phases: a spiral phase, ferromagnetism, and paramagnetism.  We analytically computed the OTOC and the entanglement entropy to reveal how the information propagates, depending on the initial phase. Here we also investigated the OTOC at different temperatures, while the entanglement entropy was considered only for the ground state as initial state. 

\begin{figure}[t]
\centerline{\includegraphics[width=1.15\linewidth]{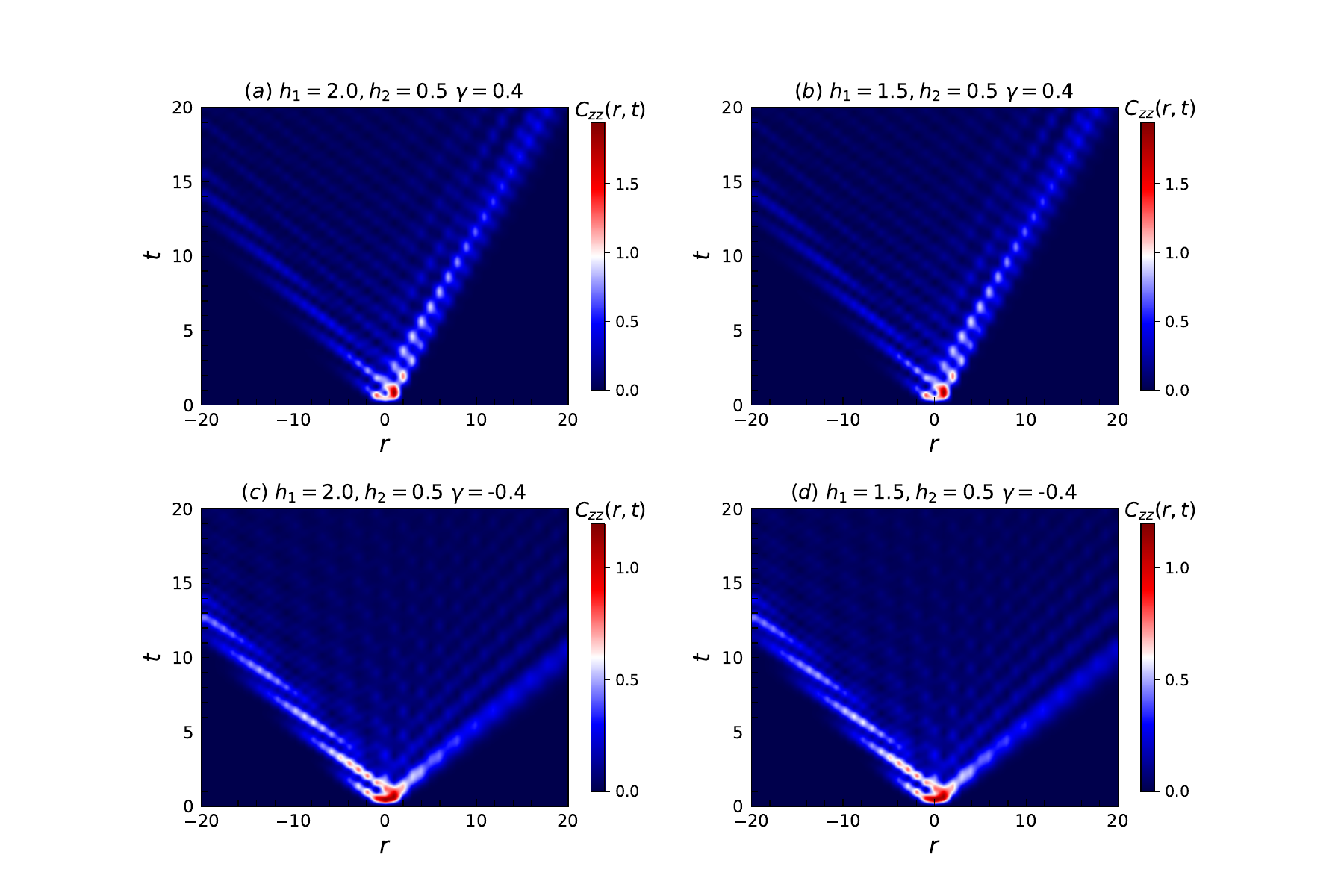}}
\caption{An example of density plot of $C_{zz}(r,t)$ versus separation, $r$, and time, $t$, under different quenches as (a) $h_1=2.0$, $h_2=0.5$, $\gamma=0.4$, (b) $h_1=1.5$, $h_2=0.5$, $\gamma=0.4$, (c) $h_1=2.0$, $h_2=0.5$, $\gamma=-0.4$ and (d) $h_1=1.5$, $h_2=0.5$, $\gamma=-0.4$. Here the size of the system is $L=100$ and $\beta=0.0$. Asymmetric  propagation is clearly visible in the figures. }
\label{Fig10}
\end{figure}

Our calculations for the butterfly velocities illustrated that the left moving butterfly velocities agree with the maximum group velocities, while the right moving ones do not. This implies that the maximum group velocity is not a strict bound for information propagation. We also found that the right butterfly velocity is larger than the maximum group velocity in the spiral phase, but smaller in the other two phases. This indicates that the operator spreading is faster in the spiral phase and slower in the other phases. Moreover, we observed that the FM phase has a smaller light cone than the other phases, which reflects the slower information propagation in this phase. We further showed that temperature does not affect the shape of the light cone.

We then investigated the effect of the $\Gamma$ interaction on the entanglement entropy of the system following a quenches across critical lines. Our results show that depending on the quench, $\Gamma$ is able to increase or decrease the value of the entanglement entropy. In addition, it can be used as a parameter to control the initial entanglement growth in the system. We demonstrated that the dynamics of the entanglement entropy can expose signals for the existence of the spiral phase. In quenches from the spiral phase, the entanglement entropy grows initially as a two-step linear growth. We also focused on the central charge within and on the boundaries of the spiral phase. We indicated that on the critical lines between the spiral phase with the FM and PM phases, the central charge is zero while within the spiral phase, it is equal to one. This is the same as the Luttinger liquid phase, revealing that the spiral phase acts like a critical region. Further studies on the dynamics of systems including $\Gamma$ interactions, especially an extension to 1D non-integrable systems as well as 2D systems would be interesting routes to follow up this work.

\acknowledgments
This work was supported by the National Science Centre (NCN, Poland) by the grant 2019/35/B/ST3/03625 (NS and HC).

\appendix\

\begin{figure}[t]
\centerline{\includegraphics[width=0.85\linewidth]{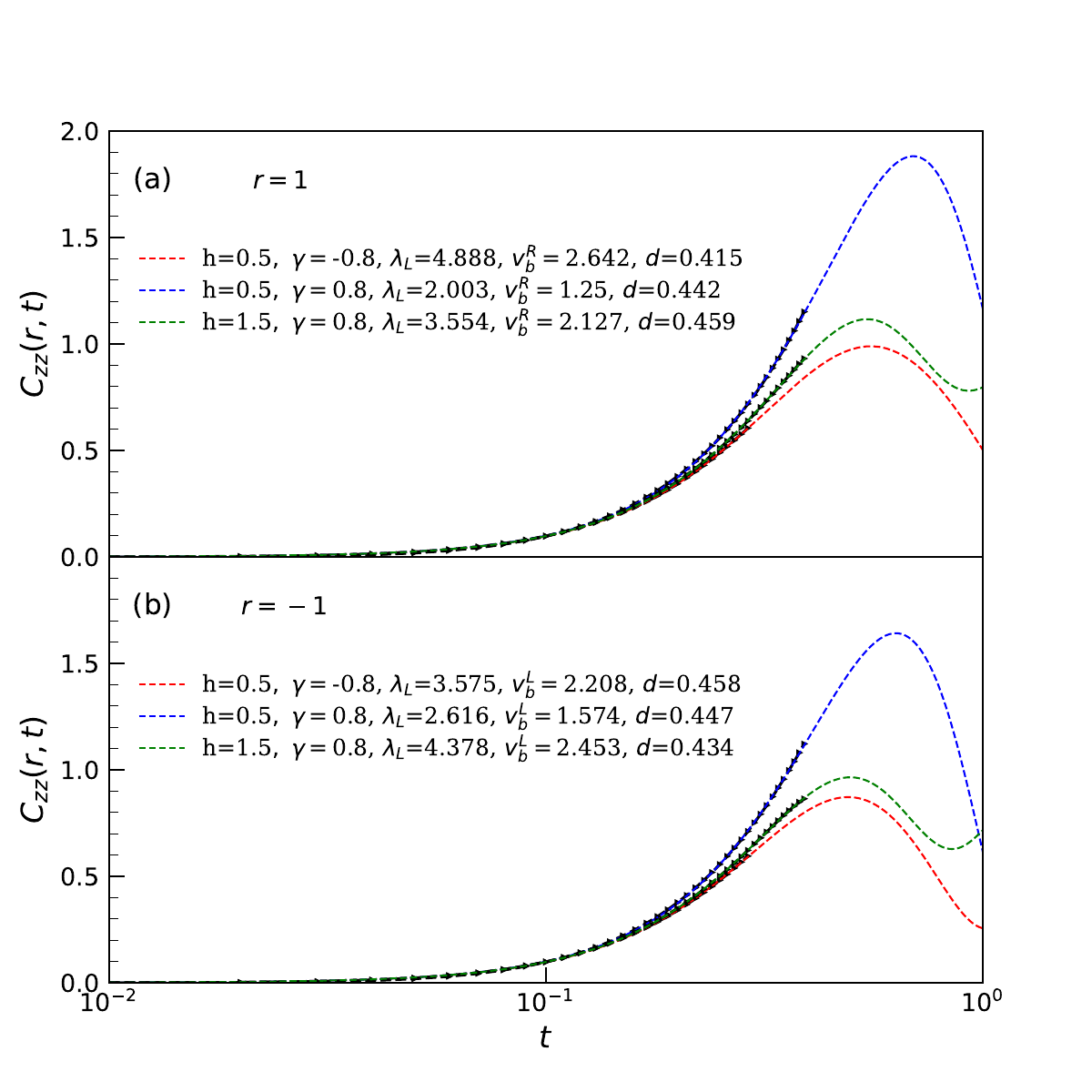}}
\caption{An example of the fitting with $L=100$, $\beta=0.0$ using eq.~(\ref{fitting}) within three phases for (a) $r=1.0$ and (b) $r=-1.0$. The black dashed lines indicate fitting functions.}
\label{Fig11}
\end{figure}

\section{OTOCs following quenches}\label{app:quenches}
Here we will consider OTOCs where the initial state is not the ground state of the time evolving Hamiltonian, i.e.~quenches. Our results show that when we do a quench in the system, the initial state does not effect how information spreads~\cite{Martyna2023g}. In contrast, the final Hamiltonian controls the different behaviour observed. As an example in Fig.~\ref{Fig10} we have plotted the density plot of $C_{zz}(r,t)$ versus $r$ and  $t$, under different quenches from the PM phase into the FM phase with $h_2=0.5$, $\gamma=0.4$ as (a) $h_1=2.0$ (b) $h_1=1.5$, and into the spiral phase with $h_2=0.5$, $\gamma=-0.4$ as (c) $h_1=2.0$, and (d) $h_1=1.5$, for a chain with size $L=100$ at $\beta=0.0$.

\begin{figure}[t]
\centerline{\includegraphics[width=1.15\linewidth]{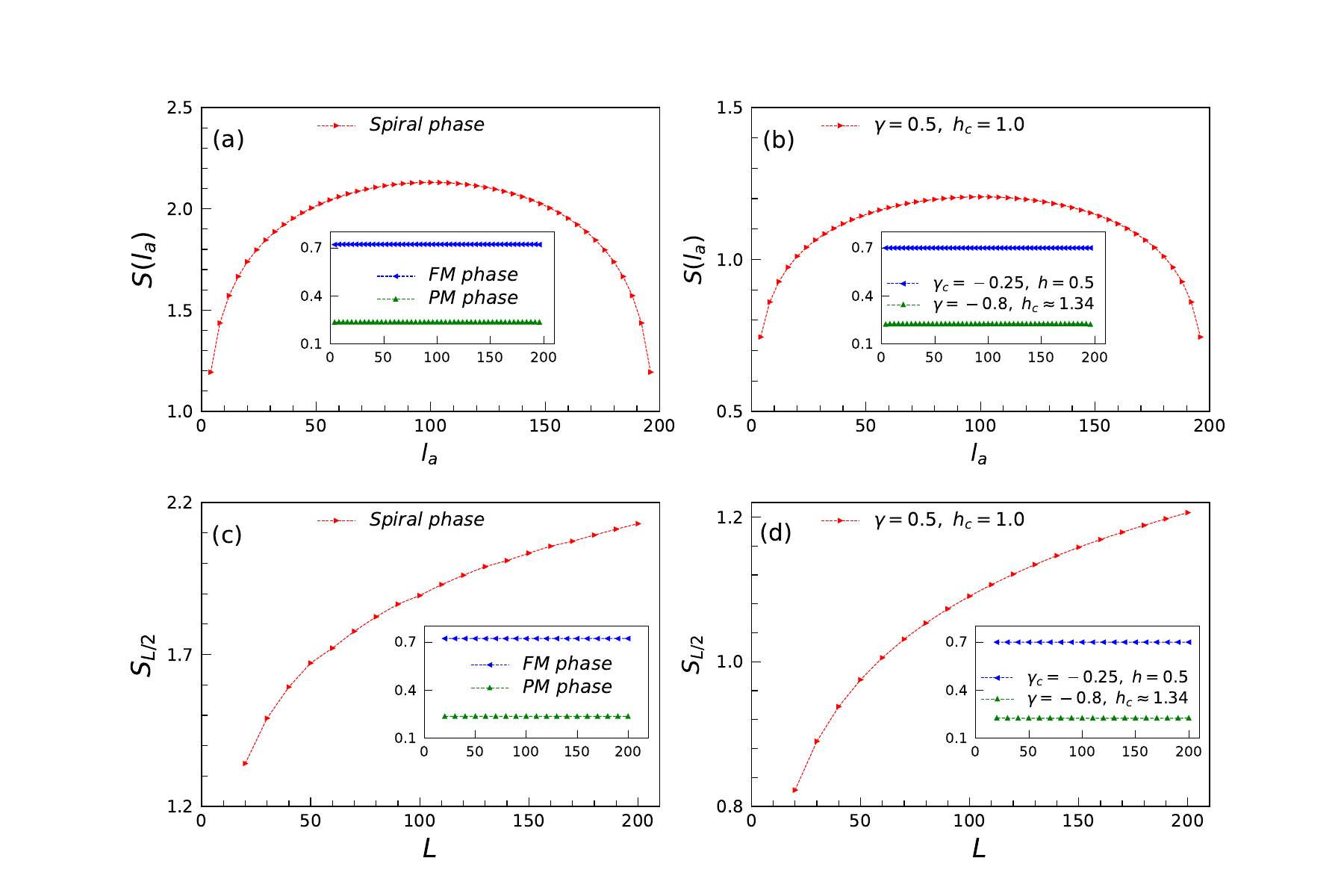}}
\caption{The entanglement entropy at the equilibrium as a function of the different subsystems $l_a$ for a fixed chain with $L=200$ (a) within the three phases and (b) on the three critical lines, and as a function of the different system sizes (c) within the three phases and (d) on the three critical lines. }
\label{Fig12}
\end{figure}

\section{Fits for Lieb-Robinson bound and the Lyapunov exponent}\label{FittingLieb}

The early time behavior and examples of the fitting procedure for the OTOC within the three phases at $\beta=0.0$ is displayed in Fig.~\ref{Fig11}. 
The fitting directly reveals different values for $v_b$, $\lambda_L$, and $d$.

\section{Equilibrium behavior of the entanglement entropy}\label{Equilibrium}

In Fig~.\ref{Fig12}(a,b), we have plotted the entanglement entropy for different subsystems where the system size is kept at $L=200$. Differences are seen only within the spiral phase and on the critical line between the FM and PM phases. Furthermore, in order to find the results for the central charge within the phases (\ref{Fig12}(c)) and on the critical lines (\ref{Fig12}(d)), we here have investigated the entanglement entropy for different system sizes. From the numerical fitting, we find the central charges 1 for the spiral phase, 1/2 for 
the critical line between the FM and PM phases, and 0 in all other phases and critical lines. 
Consequently, we claim that the spiral phase behaves critically.


\bibliography{library}

\end{document}